\documentclass[prl,onecolumn,namsmath,amssymb]{revtex4}
\usepackage{latexsym,amsmath,graphics,graphicx}

\newcommand{\otop}[1]{\overset{\circ}{#1}\,\!}

\def\be{\begin{equation}}
\def\ee{\end{equation}}
\def\beq{\begin{eqnarray}}
\def\eeq{\end{eqnarray}}
\begin{document}
\title{Theory of real space imaging of Fermi surfaces}
\author{Samir Lounis$^{1,2}$}\email{slounis@uci.edu}
\author{Peter Zahn$^3$}
\author{Alexander Weismann$^4$}
\author{Martin Wenderoth$^5$}
\author{Rainer G. Ulbrich$^5$}
\author{Ingrid Mertig$^3$}
\author{Peter H.~Dederichs$^1$}
\author{Stefan Bl\"ugel$^1$}
\affiliation{$^1$ Institut f\"ur
Festk\"orperforschung \& Institute for Advanced Simulation, Forschungszentrum J\"ulich \& JARA, D-52425 J\"ulich,
Germany}
\affiliation{$^2$ Department of Physics and Astronomy, University of California Irvine, California, 92697 USA}
\affiliation{$^3$ Institut f\"ur Physik, Martin-Luther-Universit\"at Halle-Wittenberg,
06099 Halle, Germany}
\affiliation{$^4$ Institut f\"ur Experimentelle und Angewandte Physik, Christian-Albrechts-Universit\"at zu Kiel, D-24098 Kiel, Germany}
\affiliation{$^5$ IV. Physikalisches Institut, Universit\"at G\"ottingen, 37077 G\"ottingen, Germany}
\date{\today}

\begin{abstract}
A scanning tunneling microscope can be used to visualize in real
space Fermi surfaces with buried impurities far below substrates
acting as local probes. A theory describing this feature is
developed based on the stationary phase approximation. It is
demonstrated how a Fermi surface of a material acts as a mirror
focusing electrons that scatter at hidden impurities.
\end{abstract}
\maketitle

{\bf Introduction.} The scanning tunneling microscope (STM) and spectroscopy provide a unique way of vizualising 
quantum effects from the most basic to the extremely complex ones. Among a wider range of capabilities, STM allows, for example, the investigation of electronic wave interferences after electrons 
scatter at defects on surfaces. Ring-like charge oscillations 
have been observed around Cs adatoms on Ag(111)\cite{schneider}, electrons 
waves might even be confined in corrals\cite{eigler,corral_stepanyuk,corral_szunyogh} and 
used for quantum holography\cite{moon}. These oscillations are important to understand since they 
mediate, for example, interactions between atoms sitting on surface\cite{wiebe1,wiebe2}.

In contrast to surface impurities, research on subsurface defects has 
been less intense because of the inherent experimental and theoretical 
difficulties involved in the investigation. 
Recently, a strong stream is being created towards the ultimate goal of understanding buried defects 
and their accompanying electronic waves interferences\cite{schmid,crampin,sprunger,quaas,kurnosikov,heinze}. 

In particular, we have shown that these 
interferences can be surprisingly localized and anisotropic on a real system\cite{weismann,heinrich,lounis_talk,lounis_thesis,weismann_thesis}. Using STM and 
first-principles calculations, we have demonstrated that such effects are induced by the shape 
of the Fermi surface 
of the bulk substrate, if we probe the Fermi energy in the experiment. Considering copper as a host and cobalt as
impurities, we showed that the very simple Fermi surface of copper bears very flat 
regions that 
cause, surprisingly, strong anisotropy of the screening charge distribution.  For instance,
our observations allowed us to conclude that Fermi surfaces can be
vizualised in real space with STM. Additionally, we proposed to utilize 
a buried impurity surfaces as a local probe for a nanosonar device that is
able to map buried defects and interfaces and many of their
properties, {\it e.g.} electronic or magnetic~\cite{weismann}.

This
is not unique to copper since nature is full of more complex Fermi
surfaces that could lead to unforeseen consequences. 
Besides the works cited earlier, there have been other works discussing the physics of buried 
effects. Brovko {\it et al.}\cite{stepanyuk} discuss the interesting possibility of detecting the magnetism 
of large Co-nanostructures buried below Cu(111) surface. With the nanostructure size chosen, though, it is difficult to untie and 
see a focusing effect and observe the bulk Fermi surface.
 By developing models to calculate the conductance measurable 
by STM, Avotina and co-workers\cite{avotina} performed a thorough 
investigation of the effect of Fermi surface shape. Garcia-Vidal {\it et al.}
\cite{garcia-vidal} and Reuter {\it et al.}\cite{reuter} have observed similar focusing effects in Au/Si and CoSi$_2$/Si interfaces from another 
perspective provided with 
ballistic electron emission microscopy. To understand their results, they have developed a theory based on the Keldysh formalism.

Our goal is to present a demonstration and a simple theory that takes into account the bandstructure of the host and  
the coupling of the electronic states to the defects they scatter at. Our theory is based on an analysis of Friedel oscillations around impurities by using the so-called stationary phase approximation valid for large distances 
away from the impurity.
 Discussing the effect of the Fermi surface on 
the electronic propagation goes back to the seminal work of Roth and co-workers\cite{roth} who developed a theory for the 
Ruderman-Kittel-Kasuya-Yosida (RKKY) interactions\cite{RKKY} mediated by anisotropic Fermi surfaces. This theory was behind the 
oscillation behavior of the interlayer magnetic exchange coupling\cite{gruenberg,parkin,bruno} in terms of calipers of the interlayer 
Fermi surface. Our aim is to describe the local density of states (LDOS) at few Angstroms in the vacuum above the surface 
that bears a buried single impurity. According to the model of Tersoff-Hamann~\cite{tersoff},
the LDOS is proportional to the experimental scanning tunneling spectra signal. Before discussing the effect of 
the surface we will develop the theory for a pure bulk material.

{\bf Theory of Asymptotic Behavior.} As mentioned previously, our goal
is to get the asymptotic behavior of the charge oscillations far from
the impurtiy. We use
Green functions since they allow to work with the Dyson equation. Once
known, the
space and energy resolved charge can be extracted from the Green function 
$G(\vec{r},\vec{r}'; E)$ for $\vec{r}=\vec{r}'$:
\be
 n(\vec{r};E)=-\frac{1}{\pi}\mathrm{Im}G(\vec{r},\vec{r}; E).
\ee

 The Dyson equation giving
the Green functions of an ideal host 
perturbed by an
impurity can be written as following
\beq
G^{}(\vec{r},\vec{r}'; E)&=&
\otop{G}^{}(\vec{r},\vec{r}'; E) + \int\int d\vec{r_1}d\vec{r_2}
\otop{G}^{}(\vec{r},\vec{r_1}; E) \times \nonumber \\
&\times&
t^{}(\vec{r_1},\vec{r_2}; E)
\otop{G}^{}(\vec{r_2},\vec{r}'; E),\label{G-newdef}
\eeq
where $\otop{G}$ is the Green function of the host and $t$ the t-matrix 
of the impurity being related to the potential perturbation $\Delta
V(\vec{r}) = V(\vec{r})-\otop{V}(\vec{r})$ induced by the
impurity and being given by (in formal notation) 

\beq
t&=& \Delta V \frac{1}{1-\otop{G}\Delta V}
\eeq

We use cell-centered coordinates by replacing  
$\vec{r}$ and $\vec{r}'$ by $\vec{R}+\vec{r}$ and $\vec{R}'+\vec{r}'$, 
where $\vec{R}$ and $\vec{R}'$ are lattice vectors and $\vec{r}$ and 
$\vec{r}'$ positions in the unit cell.
In the following, we assume that the impurity potential $\Delta V(\vec{r})$ 
and the related $t$-matrix $t(\vec{r},\vec{r}'; E)$ are localized in the cell 
$0$ of the impurity. The extension to a more extended perturbation is
straightforward. 

The unperturbed Green function $\otop G$ of the ideal crystal in eq.~\ref{G-newdef} can be represented by the spectral representation
\be
\otop{G}(\vec{r}+\vec{R},\vec{r}'+\vec{R}'; E)= \sum_{\nu} \frac{1}{V_B} \int d\vec{k}
\frac{\Psi_{\vec{k}\nu}(\vec{r}+\vec{R})
\Psi_{\vec{k}\nu}^*(\vec{r}'+\vec{R}')}{E+i\epsilon -E_{\vec{k}\nu}}
\quad,\label{spectral}
\ee
with $V_B$ being the volume of the Brillouin zone, $\nu$ the band
index and taking the limit $\epsilon\rightarrow +0$. 
Using additionally 
the translation symmetry of the Bloch functions $\Psi_{\vec{k}\nu}$ 
\beq
\Psi_{\vec{k}\nu}(\vec{r}+\vec{R}) &=&
 e^{i\vec{k}(\vec{r}+\vec{R})}U_{\vec{k}\nu}(\vec{r}) \ \ \mathrm{with} \ \
U_{\vec{k}\nu}(\vec{r} + \vec{R}) = U_{\vec{k}\nu}(\vec{r})
\quad .
\eeq

We can rewrite eq.~\ref{G-newdef} for the difference Green function 
$\Delta G^{}(\vec{r}+\vec{R},\vec{r}+\vec{R}; E) = G^{}(\vec{r}+\vec{R},\vec{r}+\vec{R}; E) -
\otop{G}^{}(\vec{r}+\vec{R},\vec{r}+\vec{R}; E)$ 
or $\Delta G^{\vec{R}}(\vec{r},\vec{r}; E)$, assuming the impurity at
position $R=0$, as
\beq
\Delta G^{\vec{R}}(\vec{r},\vec{r}; E) &=& 
\int\int d\vec{r}_1 d\vec{r}_2 \otop{G}(\vec{r}+\vec{R},\vec{r}_1; E) 
t(\vec{r}_1,\vec{r}_2; E) \otop{G}(\vec{r}_2,\vec{r}+\vec{R}; E)
\eeq
Using the spectral representation from eq.~\ref{spectral} we can
formulate as
\beq
\Delta G^{\vec{R}}(\vec{r},\vec{r}; E) &=& 
\sum_{\nu\nu'}\frac{1}{V_{B}^2}\int \int
d\vec{k}  d\vec{k'} \nonumber\\
&\times& \frac{e^{i\vec{k}(\vec{r}+\vec{R})}e^{-i\vec{k}'(\vec{r}+\vec{R})}}
{(E+i\epsilon - E_{\vec{k}\nu})(E+i\epsilon - E_{\vec{k}'\nu'})}
\nonumber\\ 
&\times&U_{\vec{k}\nu}(\vec{r}) \;t_{\vec{k}\vec{k'}}^{\nu\nu'}(E)\;
U^*_{\vec{k'}\nu'}(\vec{r})\label{G-newdef2}
\eeq
with the $t$-matrix elements $t_{\vec{k}\vec{k'}}^{\nu\nu'}(E)$ given by 
\beq
t_{\vec{k}\vec{k'}}^{\nu\nu'}(E)&=&\iint d\vec{r}_1d\vec{r}_2 e^{-i\vec{k}\vec{r}_1}U_{\vec{k}\nu}^{*}(\vec{r}_1) 
t(\vec{r}_1,\vec{r}_2; E)U_{\vec{k}'\nu'}(\vec{r}_2) e^{i\vec{k}'\vec{r}_2}
\eeq
where we integrate over the volume $V_0$ of the unit cell at the impurity 
site $0$. Here $t_{\vec{k}\vec{k'}}^{\nu\nu'}(E)$ is the t-matrix of the impurity 
describing the scattering process at the impurity of an incoming Bloch wave 
($\vec{k}',\nu'$) into an outgoing one with ($\vec{k},\nu$). 

In order to evaluate the difference Green function $\Delta G^{\vec{R}}$ of 
eq.~\ref{G-newdef2} at large distances $\vec{R}$ away from the impurity 
at position $R=0$, we 
analyze $\otop{G}(\vec{r}+\vec{R},\vec{r}_1; E)$ as well as 
$\otop{G}(\vec{r}_2,\vec{r}+\vec{R})$ for large distances $\vec{R}$
\beq
\otop{G}(\vec{r}+\vec{R},\vec{r}_1; E)&=&\sum_{\nu}\frac{1}{V_B} 
\int d\vec{k} \frac{e^{i\vec{k}\vec{R}}\Psi_{\vec{k}\nu}(\vec{r})
\Psi_{\vec{k}\nu}^*(\vec{r}_1)}{E+i\epsilon - E_{\vec{k}\nu}}\label{spectral2}
\eeq
To be able to evaluate the integral by the method of stationary phases, we replace the denominator by an integral over the time $t$:

\beq
\frac{1}{E+i\epsilon-E_{\vec{k}\nu}}&=&\int_0^{\infty}
\frac{dt}{i\hbar}e^{i(E+i\epsilon - E_{\vec{k}\nu})\frac{t}{\hbar}}
\eeq

The resulting double integral over $\vec{k}$ and $t$
\beq
\otop{G}(\vec{r}+\vec{R},\vec{r}_1; E)&=&\sum_{\nu}\frac{1}{V_B}
\int d\vec{k}\int_0^{\infty}\frac{dt}{i\hbar} e^{i\phi_{\nu}(\vec{k},t)}
\Psi_{\vec{k}\nu}(\vec{r})\Psi_{\vec{k}\nu}^*(\vec{r}_1)\label{spectral3}
\eeq
is dominated by a phase $\phi_{\nu}(\vec{k},t)$ varying fastly as a function 
of $\vec{k}$ and $t$
\beq
\phi_{\nu}(\vec{k},t)&=&\vec{k}\vec{R} + (E+i\epsilon - E_{\vec{k}\nu})\frac{t}{\hbar}
\eeq

Thus for large $\vec{R}$, and connected with this are large times $t$, the 
factor $e^{i\phi}$ oscillates strongly so that important contributions to the integral arize only from regions close to stationary points of $\phi(\vec{k},t)$ being given by

\be
\frac{\partial \phi}{\partial \vec{k}}= 0= \vec{R} -\frac{1}{\hbar} 
\frac{\partial E_{\vec{k}\nu}}{\partial \vec{k}}t \ \ \mathrm{and}\ \ 
 \frac{\partial \phi}{\partial t}= 0= E - E_{\vec{k}\nu}
\ee

Thus contributions to the integrals are only expected from $\vec{k}$-points 
with $E_{\vec{k}\nu}$ values close to the energy $E$ and group velocities 
$\vec{v}_{\vec{k}\nu}=\frac{\partial E_{\vec{k}\nu}}{\partial \vec{k}}$ with 
directions close to $\vec{R}$ and further times $t$ close to
$\vec{R}=\vec{v}_{\vec{k}\nu} t$. For the evaluation, we first devide
the $\vec k$-integral into a two-dimensional integral over the constant energy surface $E_{\vec{k}\nu} = \mathrm{const.}$ and a one-dimensional integral $dk_z$ perpendicular to this 
surface, with the direction $z$ given by the gradient 
$\frac{\partial E_{\vec{k}\nu}}{\partial \vec{k}}$. For the integration over 
$\vec{k}$ and $t$ we expand the phase $\phi_{\nu}(\vec{k},t)$ to second order in 
small deviations $\Delta \vec{k}_j$ and $\Delta t_j$ around these stationary 
points $\vec{k}_j$ and $t_j$

\beq
\phi_{\nu}(\vec{k}_j+\Delta \vec{k}_j,t_j + \Delta t_j)&=&
\vec{k}_j \vec{R} \nonumber\\ 
&-&\frac{1}{2}\frac{{t}_j}{\hbar}\sum_{\alpha,\beta}
\frac{\partial^2E_{\vec{k}\nu}}{\partial k_{\alpha} \partial k_{\beta}}
\bigg|_{k_j}\Delta k_{\alpha j} \Delta k_{\beta j} - \frac{1}{\hbar} 
\sum_{\alpha}\frac{\partial E}{\partial k_{\alpha}}\bigg|_{k_j} 
\Delta k_{\alpha j}\Delta t_j\label{expansion}
\eeq

In the stationary phase approximation the integral (\ref{spectral2}) can 
be evaluated analytically, provided expansion (\ref{expansion}) is used 
for the phases $\phi_{\nu}(\vec{k},t)$ and the wave functions $\Psi_{\vec{k}\nu}$ and $\Psi_{\vec{k}\nu}^*$ are replaced by the values at the stationary points 
$\vec{k}_j$. 
By determining the stationary points of all bands $\nu$ the index $j$
includes also the band index in the following derivations.
In particular, the integration over the time $\Delta t_j$ gives, 
when the integration is extended from $-\infty$ to $+\infty$:
\beq
\int_{-\infty}^{+\infty} \frac{d\Delta t_j}{\hbar} 
e^{-i\vec{v}_j\Delta \vec{k}_j \Delta t_j} &=& -\frac{2\pi i}{\hbar v_j} 
\delta(\Delta k_{zj})
\eeq
for the $z$-component of $\Delta \vec{k}_j$ in the direction of $\vec{R}$ 
coinciding with the direction of the group velocity $\vec{v}_j=\frac{dE}{\hbar d\vec{k}_j}$. For the integration over the energy plane perpendicular to $\vec{v}_{j}$ we introduce new coordinates $\Delta k_{xj}$ and 
$\Delta k_{yj}$ such that the mass tensor $\frac{\partial^2E}{\partial 
\Delta k_x \partial \Delta k_y}$ is diagonal. Using moreover the identity 
\beq
\int_{-\infty}^{\infty} dx e^{-i\theta x^2} &=& \sqrt{\frac{\pi}{|\theta|}}
e^{-i\frac{\pi}{4}\mathrm{sign}\theta}
\eeq
we obtain for the Green function (\ref{spectral3}) for very large 
$\vec{R}$ values:
\beq
\otop{G}(\vec{r}+\vec{R},\vec{r}_1; E)&\sim& -\sum_j\frac{4\pi^2i}{V_BR}
\bigg\{\bigg|\frac{\partial^2 E}{\partial k_{xj}^2}\frac{\partial^2E}
{\partial k^2_{yj}}\bigg|\bigg\}^{-\frac{1}{2}}\nonumber \\ 
&&e^{i(k_{zj}R+\varphi_j)}
\Psi_{\vec{k}_j}(\vec{r})\Psi_{\vec{k}_j}^*(\vec{r}_1)
\eeq
Here we have to sum over all critical points $k_j$ compatible with 
the direction $\vec{R}$. Thus the Green function varies asymptotically as 
$\frac{1}{R}$ (as the free electron Green function) and oscillates with a 
factor $e^{ik_{zj}R}$. In addition, there is a phase factor $\varphi_j$
\beq
\varphi_j&=&-\frac{\pi}{4}\{\mathrm{sign}(\frac{\partial^2 E}{\partial k_{xj}^2}) 
+ 
\mathrm{sign}(\frac{\partial^2 E}{\partial k_{yj}^2})
\}\label{phi}
\eeq
$\varphi_j$ is respectively equal to $-\frac{\pi}{2}$, $0$ and  $\frac{\pi}{2}$ 
when $k_{zj}$ is a maximum, a saddle point and a minimum of the surface 
constant energy. Most important is that the Green function has an amplitude 
determined by the inverse square roots of the curvatures 
$\frac{\partial^2 E}{\partial k_{xj}^2}$ and $\frac{\partial^2 E}{\partial k_{yj}^2}$ at the critical point $k_j$. In the above derivation, we 
have implicitly assumed that the contribution from the previously mentioned second derivatives does not vanish at the critical point $\vec{k}_j$ which in 
general is realized. However, if vanishing derivatives occur, we speak
of 
a higher order critical point, which results in an even slower
decrease of the Green function than $\frac{1}{R}$. Since such critical points are very important for STM observations, leading for subsurface impurities to strong intensity 
``spots'' in certain directions, we will discuss these anomalies in the 
upcoming text.

For the evaluation of the difference Green function $\Delta G(\vec{r}+\vec{R},\vec{r}+\vec{R};E)$ of eq.~\ref{G-newdef2} we need the 
asymptotic expansion for both $\otop{G}(\vec{r}+\vec{R},\vec{r}_1;E)$ as 
given by eq.~\ref{phi} and the analogous one for $\otop{G}(\vec{r}_2,\vec{r}+\vec{R}; E)$. Note that the latter Green function 
describes the propagation from the cell $\vec{R}$ to the impurity cell, while the former Green function describes the back propagation from the impurity to position $\vec{R}$. Therefore in the expression for 
$\otop{G}(\vec{r}_2,\vec{r}+\vec{R}; E)$, in the equation analogous to eq.~\ref{spectral2} a factor $e^{-i\vec{k}\vec{R}}$ enters, and the role of 
$\vec{r}$ and $\vec{r}_2$ (replacing $\vec{r}_1$) has to be exchanged. Analogously, the critical points $\vec{k}_{j'}$ of $\phi_{\nu}(\vec{k},t)$ for $\otop{G}(\vec{r}_2,\vec{r}+\vec{R};E)$ are, up to a minus sign, identical with the one for $\otop{G}(\vec{r}+\vec{R},\vec{r}_1;E)$: $\vec{k}_{j'} =-\vec{k}_j$. 
Taking all this together, we obtain for the difference Green function
\beq
\Delta G(\vec{r}+\vec{R},\vec{r}+\vec{R}; E) &=&
-\sum_{jj'}
\frac{16\pi^4}{V_B^2}\Psi_{\vec{k}_j}(\vec{r})\Psi_{\vec{k}_{j'}}^*
(\vec{r})\nonumber\\
&\times&t_{\vec{k}_j\vec{k}_{j'}}(E) \frac{e^{i(k_{zj}-k_{zj'})R+i(\varphi_j+\varphi_{j'})}}
{R^2}\nonumber\\
&\times&\bigg|
\frac{\partial^2E}{\partial k_{xj}^2}
\cdot 
\frac{\partial^2E}{\partial k_{yj}^2}
\cdot
\frac{\partial^2E}{\partial k_{xj'}^2}
\cdot  
\frac{\partial^2E}{\partial k_{yj'}^2}
\bigg|^{-\frac{1}{2}}\label{18}
\eeq

For the evaluation of the energy dependent change of the charge density
$\Delta n(\vec{r}+\vec{R}; E)$ we have to take the imaginary part of 
(\ref{18}). Assuming only one pair of stationary points $j$ and $j'$, this leads to 
\beq
\Delta n(\vec{r}+\vec{R}; E)&=& -\frac{16 \pi^2}{V_{B}}
|\Psi_{\vec{k}_j}(\vec r)|^2 |t_{\vec{k}_j-\vec{k}_{j}}|\nonumber \\
&\times& \frac{\sin{(2k_{zj}R+2\varphi_j+\delta_{\vec{k}_j})}}
{R^2\bigg|\frac{\partial^2E}{\partial k_{xj}^2} 
\frac{\partial^2E}{\partial k_{yj}^2}\bigg|}\label{20}
\eeq
where $\delta_{\vec{k}_j}$ is the phase of the $t$-matrix
$t_{\vec{k}_j\,-\vec{k}_{j}}$. In order to get the change $\Delta
n(\vec{r}+\vec{R})$ of the charge density, we have to integrate the
strongly varying part over the energy of the occupied states
\beq
-\int^{E_F} dE \sin{(2k_{zj}R+2\varphi_j+\delta_{\vec{k}\nu})}&=&
-\hbar v_j \int^{k^{E_F}} dk_{zj} \sin{(2k_{zj}R+2\varphi_j+\delta_{\vec{k}_j})}\nonumber\\
=\frac{\hbar v_j^{E_F}}{2R}\cos{(2k_{zj}^{E_F}R+2\varphi_j+\delta_{\vec{k}_j})}
\eeq
and can replace the energy in all other parts by $E_F$. Thus $\Delta n(\vec{r}+\vec{R})$ is given by
\beq
\Delta n(\vec{r}+\vec{R}) &=& \frac{16 \pi^2 \hbar v_j}{V_B}
|\Psi_{\vec{k}_j}(\vec{r})|^2|t_{\vec{k}_j\,-\vec{k}_{j}}| 
\frac{\cos{(2k_{zj}R+2\varphi_j+\delta_{\vec{k}_j})}}
{R^3\bigg|\frac{\partial^2 E}{\partial k_{xj}^2}
\frac{\partial^2 E}{\partial k_{yz}^2}\bigg|}\label{22}
\eeq

At this stage, we can discuss the physical implications of the
previous formula. We see from the last equation
eq.~(\ref{22}) that the denominator is a crucial
factor. If the denominator is very small meaning that the constant
energy  surface has a flat region, big values of the 
charge density are obtained from this $\vec{k}$-region, leading to a strong
focusing of intensity in space region as determined by the group
velocity $\vec{v}_j$. In other words, the curvature of
the constant energy surface defines the focusing of the charge.

{\bf Few ab-initio results.} Let us now go back to the experimentally investigated case: a Co
impurity sitting below the surface of Cu(111)\cite{weismann}.
We use the  Full-potential Korringa-Kohn-Rostoker Green function (KKR-GF) method~\cite{kkr} which  is
ideal for investigating impurity problems in real space
exactly in the geometry given by 
the system of interest. For instance, 
we calculated the space-resolved local density of states (LDOS) in the 
vacuum region integrated over a small energy region around $E_F$ (from $E_F - 0.136~\mathrm{eV}$ to $E_F$) 
at  $\approx$ 6.1 \AA~above Cu(111) surface and at $\approx$ 3.5 \AA~above
Cu(001) surface. 
Below the Cu(111) surface we have considered a Co impurity at
the 6$^{th}$ and at the 3$^{rd}$ layer below the 
surface, and below the Cu(001) we consider an impurity at the $8^{th}$ 
layer below the surface.
\begin{figure}[ht!]
\begin{center}
\includegraphics*[width=.5\linewidth,angle=270]{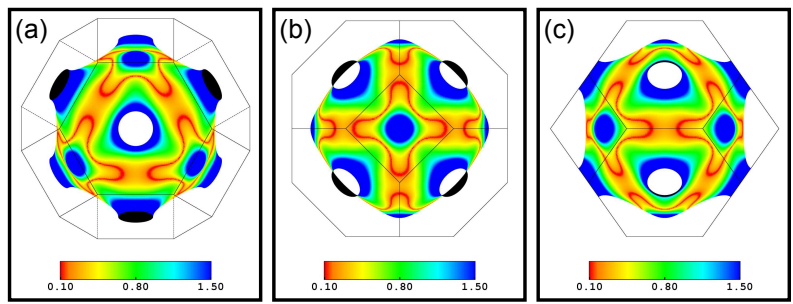}
\end{center}
\caption{Fermi surface of Copper represented along three directions:
  (111) is shown in (a), (001) in (b) and (110) in (c). The inverse
  mass tensor corresponding to the denominator of
eq.~(\ref{18}) is represented by the color in units of the inverse
electron mass. Small values represented in red lead to high intensities of
the charge variation.}
\label{compilation2}
\end{figure}

\begin{figure}[ht!]
\begin{center}
\includegraphics*[width=1.\linewidth]{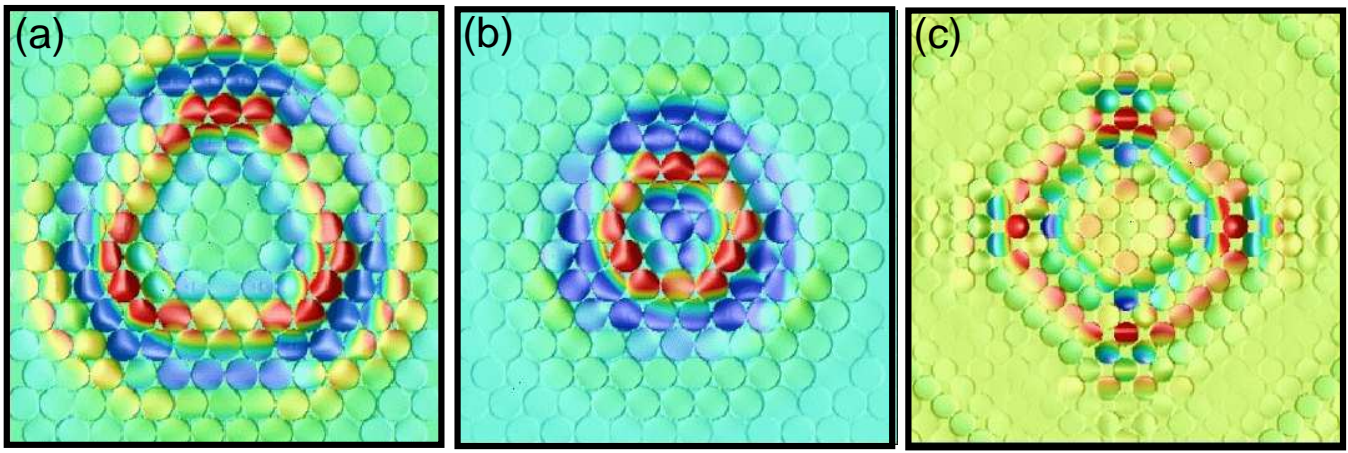}
\end{center}
\caption{(a) and (b): Impurity induced charge density around
$E_F$  at a height of $\approx$ 6.1 {\AA} above the Cu(111) surface
with an Co impurity sitting in the 6$^{th}$ layer (a) and in
the 3$^{rd}$ layer (b) below the surface; (c): The case of an impurity buried at 8 layers below 
a Cu(001) surface is shown.
Red/blue color means enhancement/reduction of the local density of states
at $E_F$.}
\label{compilation}
\end{figure}

\begin{figure}[ht!]
\begin{center}
\includegraphics*[width=\linewidth,angle=0]{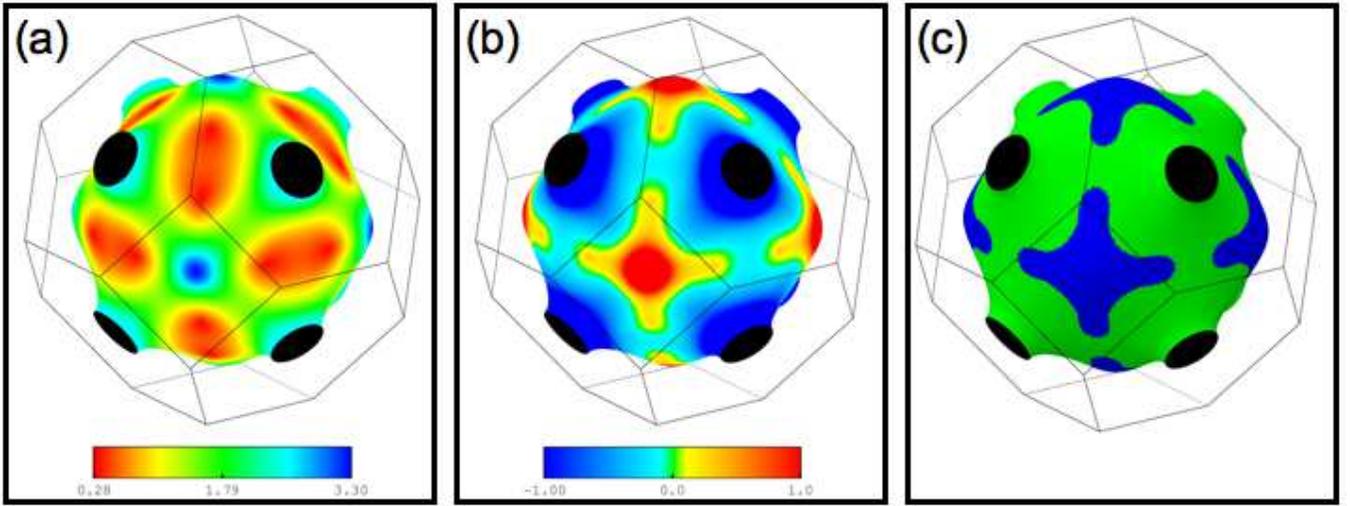}
\end{center}
\caption{
  Fermi surface of Copper showing the diagonal elements of the
  inverse mass tensor in x and y direction in (a) and
  (b), respectively. 
  In panel (c) the Fermi surface of copper is colored following the
  curvature of the Cu-Fermi surface, which determines the amplitude 
  and the phase of the oscillations in eq.~\ref{phi} and eq.~\ref{20}. 
 Blue indicates an inward curvature like the curvature of an ellipsoid, green means an outward curvature. At the boundary line between the green and blue areas the Gaussian curvature vanishes, indicating higher order critical behaviour as in Eq.~\ref{highorder}.
} 
\label{compilation3}
\end{figure}

 The Cu Fermi
surface is rather spherical apart from the band gaps in the
$L$-directions; flat areas with strongly reduced curvature are
present in the (110) directions enclosed by the two (111)-necks
and two elevations in the (001) directions (see
Fig.~\ref{compilation2}). These flat regions are 
represented along the three directions (111), (001) and (110) in red in Fig~\ref{compilation2}. The colour scale
on the Fermi surface represent the strength of the inverse mass
tensor (denominator of eq.~(\ref{18})) which
measures the flatness. This explains the anisotropic charge
ripples observed experimentally and calculated from
first-principles~\cite{weismann}. Interestingly, along the (111) direction, the
neck of the Fermi surface defines a forbidden region where no
electrons can scatter explaining the flat region in the charge density
changes at the center of
Fig.~\ref{compilation}(a) and Fig.~\ref{compilation}(b). The latter 
figures are different since they are produced by an impurity buried 
at different distances from the surface: in (a) it is at 6 layers below 
the surface while in (b) it sits much closer to the surface 
at the $3^{rd}$ layer under the surface. It is interesting to note
that when the atom is closer to the surface, the anisotropy of the
oscillations seems to loose in intensity which is induced by the
stronger scattering of the surface state electrons present on the
(111) surface of copper. Those surface state electrons are associated with a nearly isotropic two-dimensional circular Fermi surface. In Fig.~\ref{compilation}(c), parts of 
the Fermi surface along the (001) direction (see Fig.~\ref{compilation2}(b)) is probed. This is perfomed by 
assuming a buried impurity in a Cu(001) sample. To improve the visualization of the curvature change observed on the Fermi surface of copper, the Fermi surface is colored with blue and green in  
Fig.~\ref{compilation3}(c). Here, the two regions with opposite
curvatures 
are obvious and are sepated by a region with a low curvature that induces the focusing effect.

\begin{figure}[ht!]
\begin{center}
\includegraphics*[width=1.\linewidth]{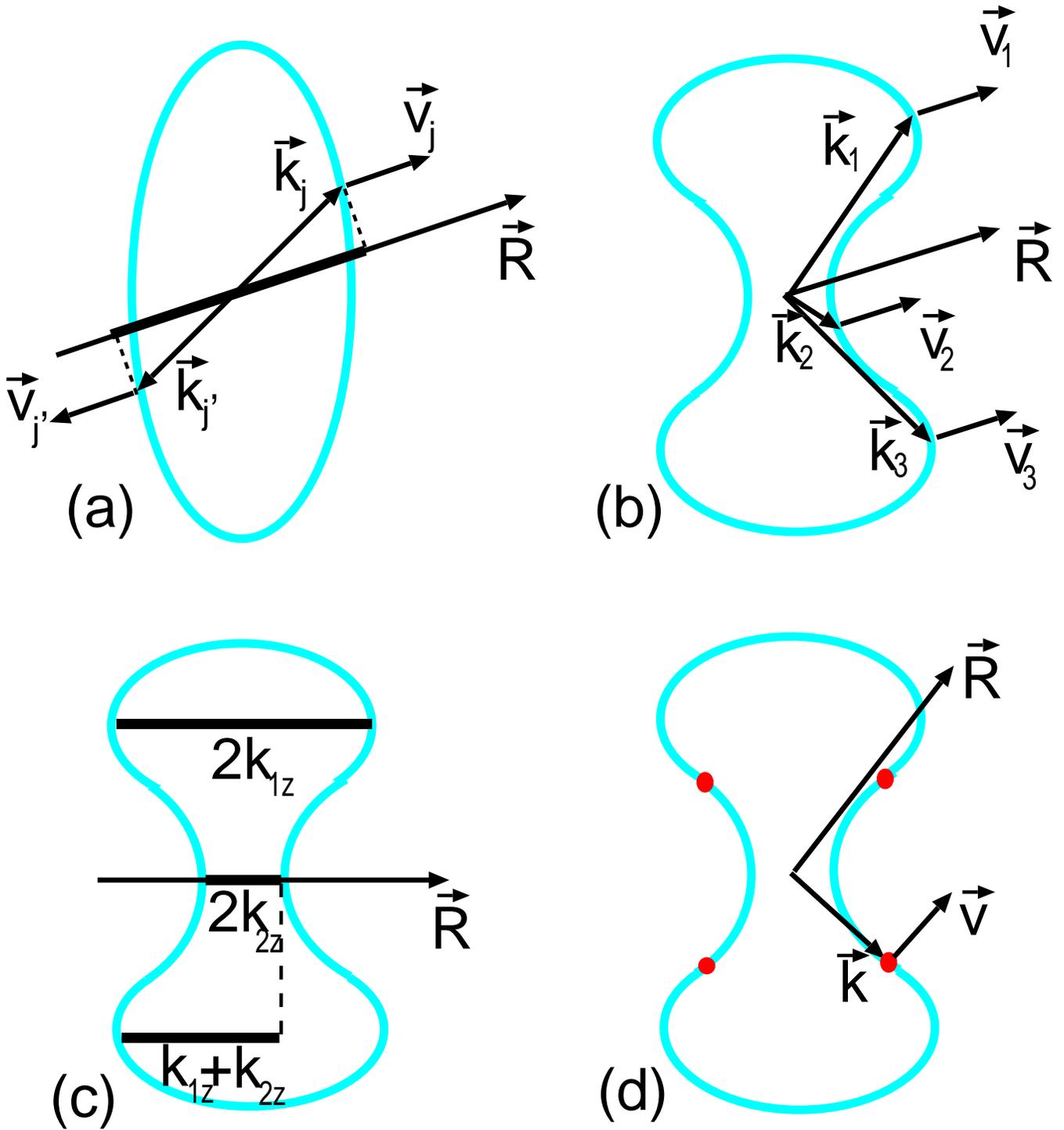}
\end{center}
\caption{Examples of Fermi contours for illustrative purposes. (a)
  show an ellipsoidal Fermi surface, which for a given $\vec{R}$ 
  has one critical point with a group velocity $\vec{v}$ parallel to
  $\vec{R}$. In (a) is shown a thick line which length defines the
  oscillation period. The Fermi surface sections (b), (c) and  
(d) are similar to the ``dog's bone'' of copper Fermi surface. For an
  oblique orientation of $\vec{R}$ with respect to
  the contour's long axis represented in (b), 
  several critical points
  contribute to the oscillatory behavior. In (c), however, $\vec{R}$ is
  parallel to the short axis, which reduces the number of critical
  points. In (d), red disques show possible
  inflection points corresponding to higher order critical points
  discussed in the text. One possible direction of $\vec R$ probing
  this region is shown.
} 
\label{fermi_contours}
\end{figure}

{\bf{Multiple critical points.}} While the formula (\ref{phi}) and (\ref{18}) 
are valid for an arbitrary number of critical points, we have analyzed
in the last section only the results for one pair of critical point. If for a
given $\vec{R}$-value there exist more critical points ($\vec{k}_1 ,
\vec{k}_2, ...$)  
on the Fermi surface with group velocities $\vec{v}_1, \vec{v}_2, ...$ parallel 
to $\vec{R}$, than the Green function at $E_F$ as well as the charge density 
$\Delta n(\vec{r}; E_F)$ exhibit several oscillation periods as a
function of R. For instance, 
for two points $\vec{k}_1$  and $\vec{k}_2$ with $\vec{v}_1$ and $\vec{v}_2$ parallel to $\vec{R}$, 
there are oscillation periods determined by the projections $k_{z1}$ and 
$k_{z2}$ on the direction of $\vec{R}$(see Fig.~\ref{fermi_contours}d). Due to the double sum over $j$, $j'$ 
in (\ref{18}) and (\ref{20})  the charge densities $\Delta n(\vec{r};E)$ and $\Delta n(\vec{r})$ than show three periods being determined by the $k_z$-values $2k_{z1}$, $2k_{z2}$ and $k_{z1}+k_{z2}$. The amplitude 
of these oscillations are determined by the curvatures at these $k$-points as 
well as the wavefunctions $\Psi_{\vec{k}_j}(\vec{r})$ and the $t$-matrix elements $t_{\vec{k}_j\vec{k}_{j'}}(E)$. It is easy to show that for $N$ critical points the number of periods is $\frac{N(N+1)}{2}$. Sometimes, {\it e.g.} for 
symmetry reasons, some of these periods can be the same, {\it e.g.} 
$\vec{k}_1$ and $\vec{k}_2$ can be different, but might have the same 
z-component, so that only one period $k_{z1}=k_{z2}$ exists.

The behavior is illustrated in Fig.~\ref{fermi_contours} for two Fermi
surface sections. Fig.~\ref{fermi_contours}(a) shows an ellipsoidal
Fermi surface, which for each given $\vec{R}$ vector has only one pair
of critical point 
$\vec{k}$ with group velocity $\vec{v}//\vec{R}$. Also the two vectors $\vec{k}_j$ and $\vec{k}_{j'}=-\vec{k}_j$ are shown, the projection of which on the direction $\vec{R}$ gives the diameter of the Fermi surface 
(thick line) which determines the oscillation period along the direction $\vec{R}$.

Fig.~\ref{fermi_contours}(b)-(d) show a more complicated Fermi
surface, resembling the ``dog bone'' of the Cu Fermi surface. For the
direction $\vec{R}$ shown in Fig.~\ref{fermi_contours}(b), three
different $\vec{k}_j$  
points with $\vec{v}_j\parallel\vec{R}$ exists, leading to a total of 6 different 
periods ($2k_{1z}, 2k_{2z}, 2k_{3z}, k_{1z}+k_{2z}, k_{1z}+k_{3z}, k_{2z}+k_{3z}$) 
for the charge density. On the other hand, if $\vec{R}$ is
perpendicular to the main axis as in Fig.~\ref{fermi_contours}(c),
then $\vec{k_{z1}}=\vec{k_{z3}}$  
and only three periods exist (indicated by the thick lines) while if 
$\vec{R}$ points along the main axis, there is only one solution. Thus for 
a given Fermi surface the situation can be quite complex.

{\bf{Higher order critical points.}} If one of the second derivatives 
$\frac{\partial^2E}{\partial k_x^2}$ or $\frac{\partial^2E}{\partial k_y^2}$ 
in eq.~\ref{18} or \ref{20}, \ref{22} vanishes, than the Green function expression and the charge density diverge, meaning that asymptotically these quantities decrease even with a smaller exponent than $\frac{1}{R}$, 
$\frac{1}{R^2}$ and $\frac{1}{R^3}$ respectively. We consider here
four such cases, dropping $j$ as index of of $k_j$ and $t_j$:

a)$\frac{\partial^2 E}{\partial k_x^2} = 0$, but $\frac{\partial^2 E}{\partial k_y^2} \neq 0$. In this case, we expand the phase factor of eq.~\ref{expansion} for $\Delta k_x$ up to $(\Delta k_x)^3$:
\beq
\phi(\vec{k}+\Delta{\vec{k}},t+\Delta t)&=&
kR - 
\frac{1}{6}\frac{t}{\hbar}\frac{\partial^3E}{\partial k_x^3}\Delta k_x^3 -
\frac{1}{2}\frac{t}{\hbar}\frac{\partial^2E}{\partial k_y^2}\Delta k_y^2 -
-\frac{1}{\hbar} v \Delta k_z \Delta t
\quad .
\eeq

The integration over $\Delta k_x$, $\Delta k_y$, and $\Delta k_z$, as
well as the $t$-integration can then be performed, leading to a Green
function: 
\be
G(\vec{r}+\vec{R},\vec{r}_1; E_F)\sim
 \frac{1}{(R|\frac{\partial^3E}{\partial k_x^3}|)^{\frac{1}{3}}}
\frac{1}{(R|\frac{\partial^2E}{\partial k_y^2}|)^{\frac{1}{2}}}
\sim \frac{1}{R^{\frac{5}{6}}}\label{highorder}
\ee
Thus the decay for larger distances is slightly slower than $\frac{1}{R}$. 
The charge density at $E_F$, $\Delta n(\vec{r}+\vec{R}; E_F)$ varies 
then as $\frac{1}{R^{\frac{5}{3}}}$ (instead of $\frac{1}{R^2}$),
while the total charge $\Delta n(\vec{r}+\vec{R})$ varies as
$\frac{1}{R^{\frac{8}{3}}}$ instead of the familiar $\frac{1}{R^3}$ of
typical Friedel oscillations. 
For the case of copper, we observe such a situation as shown in
  Fig.~\ref{compilation3}(a) and (b): Here the color scale on the
  Fermi surface represents the inverse masse tensors strength along  
the $x$ and $y$ directions that are defined as diagonal elements of
the inverse mass tensor of eq.~\ref{phi}. 
The directions x and y are chosen this way that 
$\frac{\partial^2E}{\partial k_x^2}$ is always larger than the
y-component. 
It is interesting to note that  
in Fig.~\ref{compilation3}(a), $\frac{\partial^2E}{\partial k_x^2}$ is
always positive for Cu, whereas
 $\frac{\partial^2E}{\partial k_y^2}$ shown in
Fig.~\ref{compilation3}(b) changes
sign and is equal to zero on the green line. 
The case considered here corresponds to the border line between the
green and blue areas in fig.~\ref{compilation3}(c)
where one of the second derivatives changes sign.

b) In case that $\frac{\partial^2 E}{\partial k_x^2} =\frac{\partial^2
  E}{\partial k_y^2} = 0$, the charge density decreases again
slower. Assuming that the third order derivatives have radial
symmetry, the difference of the Green function varies proportional to
$\frac{1}{R^{\frac{2}{3}}}$, 
$\Delta n(\vec{R}; E_F)$ decays as $\frac{1}{R^{\frac{4}{3}}}$,
 and the total charge density change drops as
 $\frac{1}{R^{\frac{7}{3}}}$. 
Such a case could be induced by the inflection points represented as a
red disk spot in Fig.~\ref{fermi_contours}(d).  

c) We consider now a case where the energy $E(k_x,k_y)$ is constant along a 
$k_y$-line of length $l$ perpendicular to $\vec{R}$, with constant group velocity pointing along $\vec{R}$. Then $\Delta G \propto \frac{1}{\big(R\big|\frac{\partial^2 E}{\partial k_x^2}\big|\big)^{\frac{1}{2}}}\cdot l$ and 
$\Delta n(\vec{R};E_F)\propto\frac{l^2}{R}$, $\Delta n(\vec{R})\propto\frac{l^2}{R^2}$. 

The decrease corresponds exactly to Friedel oscillations in two dimensions ($x$ and $z$), the third dimension $k_y$ gives just a constant factor $l$, 
respectively $l^2$ for the charge density. Thus the decrease is very slow.

d) Finally, we consider an extreme case, that $E(k_x,k_y)$ is constant
on a whole plane with edge length $l$ representing a perfectly flat part of the Fermi surface. Then the $k_x$ and $k_y$ integrations give just a factor of $l$ each, such that 
$\otop{G} \propto l^2$ and $\Delta n(\vec{R}; E_F)\propto l^4$ and $\Delta n(\vec{R})\propto\frac{l^4}{R}$. Thus the Green function $\Delta G$ and the charge density have a purely oscillatory behavior with no decrease in R, while the Friedel oscillations decrease only as $\frac{1}{R}$. This behavior is typical for a one dimensional system, and represents the slowest possible decrease. Most important is that the amplitude varies with fourth power of the length $l$. Thus planar pieces of the Fermi surface lead to particular slowly decreasing oscillations, and this effect increases strongly with the size of the platelet.

{\bf Role of the surface.} Here we would like to comment on the role of the surface.  
Up to now we have assumed a bulk system. However, as mentioned previously, 
the quantity of interest for the interpretation of STM experiments
is the LDOS inside the vacuum at a certain distance $z$ from the
surface. According to the theory of Tersoff and Harmann\cite{tersoff} this
distance corresponds to a position in the center of the STM tip so
that tunneling parameters (bias Voltage, setpoint current) as
well as the tip radius have an impact on $z$. There are different ways to 
implement the surface into the model. For simplicity, we will work with 
the spectral representation of the host Green function $\otop{G}$ from
eq.~(\ref{spectral}).

The tunneling current has contributions of multiple states, which
decay differently into the vacuum. If we assume full
translational invariance parallel to the surface, the parallel
component of wave vector $k_{\parallel}$ is conserved. Inside the
vacuum the electrons obey the Schr\"odinger equation
of a free-electron: \beq
\frac{\hbar^2}{2m}(k_{\parallel}^2-\kappa^2)=\Delta E-\Phi \eeq
Here $\Phi$ is the work function of the material and $\Delta
E=E-\mu$ the electron energy with respect to the chemical
potential. From this a $k_{\parallel}$-dependent decay constant
$\kappa$ can be derived:\beq
\kappa(k_{\parallel})=\sqrt{\frac{2m}{\hbar^2}(\Phi-\Delta
E)+k_{\parallel}^2} \label{kappa_kpar} \eeq The above expression
shows that states with high $k_{\parallel}$-values have a higher
$\kappa$ and therefore decay faster into the vacuum. Consequently
the STM is more sensitive to states near the center of the surface
Brillouin-zone with small $k_\parallel$, while short-wavelength
contributions to the LDOS oscillations, corresponding to large
$k_\parallel$, are suppressed.

In the vacuum, the wave-function of state $\Psi_{\vec k}$ is then:
\beq
\Psi(\vec{r_\parallel},z)\propto
\exp{\left(i\vec{k}_{\parallel}\vec{r_\parallel}\right)}
\cdot\exp{\left(-z\sqrt{\vec{k}_{\parallel}^2+\kappa_0^2}\right)}
\quad,
\label{dec_vac}\eeq 
where we defined
$\kappa_0\equiv\kappa(k_{\parallel}=0)=
\sqrt{2m(\Phi-\Delta E)}/\hbar$. In order to get an idea how this effect
influences the observed patterns we perform a Taylor approximation
of eq.~\ref{dec_vac} up to first order in $k_{\parallel}^2$:
\beq
\sqrt{k_{\parallel}^2+\kappa_0^2}\approx\kappa_0+
\frac{k_{\parallel}^2}{2\kappa_0}
\label{taylor}  
\quad .
\eeq 
This approximation is valid for $k_{\parallel}\ll\kappa_0$. For
the case of copper this is a good treatment of states having
$k_{\parallel}<1.1 \AA^{-1}$ while states having higher
$k_{\parallel}$ values are over-suppressed 
. Inserting
\ref{taylor} in \ref{dec_vac} gives a simple expression:
\beq
\exp{(-\kappa z)}\approx \exp{(-\kappa_0 z)}\cdot
\exp{\left(-\frac{k_{\parallel}^2z}{2\kappa_0}\right)}
\quad .
\label{gauss}\eeq 
This is, apart from a general attenuation of the
wave functions amplitude by $\exp{(-\kappa_0z)}$, a Gaussian with a
width of $\sigma_k^2=\kappa_0/z$. This implies that one
can relate the wave functions from a smaller distance $z_1$ to
greater distances $z_2$ by a convolution (symbol $\ast$) with a
Gaussian of width $\sigma_z^2=\kappa_0/(z_2-z_1)$: 
\beq
\Psi(\vec{r}_\parallel,z_2)=
\exp{\left(-\kappa_0(z_2-z_1)\right)}\cdot\Psi(\vec{r}_\parallel,z_1)\ast
\exp{\left(-\frac{\vec{k}_{\parallel}^2}{2\sigma_z^2}\right)}
\quad .
\eeq 
This is a very helpful expression as the effect of the
tip-sample distance $z$ can be described by a simple Gaussian
filtering. If we increase $z$ either by choosing tunneling
conditions, where the tip is at a larger distance from the surface
or by using a blunter tip, the wave functions probed by the tip
will be increasingly smeared out. If the approximation in eq.~\ref{gauss}
is not valid, the convolution has to be performed by a Fourier
transform of the last term 
in eq.~\ref{dec_vac}, but the whole effect can still be
understood as a kind of smoothing filter. This convolution can
also be applied to any superposition of wave functions as well as
to the Green functions which would then decay similarly.\par

In other words, one has to be carefull since additionally to the host Fermi 
surface the vacuum tunneling modifies and preselect
interferences. Interferences created in  
the bulk can be different from those measured with the STM above the
surface of the material.

{\bf Conclusion.} 
To conclude, as the Fermi-surfaces of most materials
deviate strongly from a spherical shape, the corresponding propagation
of electrons is anisotropic and could reveal new
effects in different materials present in nature. For instance,
the combination of buried impurities and a scanning tunneling
microscope could be used as a nanosonar
to investigate the interior of materials. We developed and presented a
theory behind the focusing effect of electronic wave
oscillations. Additionally, the effect of different kind of Fermi
contours' critical points are discussed and the consequences for the
decay of charge density oscillations are highlighted.

\section{A\lowercase{cknowledgements}}
This work was supported by the ESF EUROCORES Programme SONS under
contract N.\ ERAS-CT-2003-980409, the Deutsche Forschungsgemeinschaft Priority Programme SPP1153 and the Deutsche Forschungsgemeinschaft Collaborative Research Centre SFB602.
S.L. gratefully acknowledges support by the Alexander von Humboldt Foundation through
the Feodor Lynen Program and wishes to thank D. L. Mills for discussions and hospitality.

\end{document}